\documentclass[traditabstract]{aa}
\usepackage{txfonts}
\usepackage{color}
\usepackage{graphicx}
\usepackage{natbib}
\bibpunct{(}{)}{;}{a}{}{,} 

\usepackage[normalem]{ulem} 

\title{Impact of local structure on the cosmic radio dipole}

 \author{Matthias Rubart$^1$\thanks{\tt matthiasr at physik dot
uni-bielefeld dot de}, David Bacon$^2$ \thanks{\tt david dot bacon at port dot ac dot uk},
   		Dominik J. Schwarz$^1$\thanks{\tt dschwarz at physik dot
uni-bielefeld dot de}}

\institute{$^1$Fakult\"at f\"ur Physik, Universit\"at Bielefeld, Postfach 100131, 33501 Bielefeld, Germany\\
$^2$Institute of Cosmology and Gravitation, University of Portsmouth, Burnaby Road, Portsmouth PO1 3FX, 
United Kingdom}

\abstract{
We investigate the contribution that a local over- or under-density can have on linear cosmic dipole estimations. 
We focus here on radio surveys, such as the NRAO VLA Sky Survey (NVSS), and forthcoming surveys such as those with the LOw Frequency ARray (LOFAR), the Australian Square Kilometre Array Pathfinder (ASKAP) and the Square Kilometre Array (SKA). The NVSS has already been used to estimate the cosmic radio dipole; it was shown recently that this radio dipole amplitude is larger than expected from a purely kinematic effect, assuming the velocity inferred from the dipole of the cosmic microwave background. We show here that a significant contribution to this excess could come from a local void or similar structure. In contrast to the kinetic contribution to the radio dipole, the structure dipole depends on the flux threshold of the survey and the wave band, which opens an opportunity to distinguish the two contributions.}


\titlerunning{Local structure dipole}
\authorrunning{Matthias Rubart, David Bacon \& Dominik J. Schwarz}

\begin{document}
\maketitle

\section{Introduction}

In recent years,  the dipole anisotropy in radio surveys, such as the NVSS catalogue \citep{NVSS}, has been 
investigated (e.g. \citet{BW02}, \citet{Singal11}, \citet{Gibelyou}, \citet{firstpaper} and \citet{Jain13}). It appears 
that the cosmic radio dipole has a similar direction to the one found in the Cosmic Microwave 
Background (CMB), but with a significantly higher amplitude (by a factor of three to four, based on 
different estimators and surveys \citep{Singal11, firstpaper}). 
In this work we investigate one possible effect which can increase the dipole amplitude observed in radio 
surveys, with respect to the CMB dipole. 

There have recently been studies (e.g. \citet{Void1} and \citet{Void2}) which claim that the local 
universe (i.e. on scales of 300 Mpc) has an atypically low density of galaxies. If we do live in 
such a region, what would we expect to see regarding the observed cosmic radio dipole? We are unlikely to be
living in the very centre of such a void, so  there will be some offset distance between us 
and the centre of the void, which we call $\vec r_{\rm v}$. If we imagine a sphere around 
the observer (in our case the Local Group), with a radius $R_{\rm o}$ greater 
than the void radius $R_{\rm v}$, we will expect to see more galaxies in one direction than in the other. 

It is likely that the Local Group moves towards the direction where we see more
galaxies, due to their gravitational pull. This direction has been determined to be 
$(l,b)$ = ($276^\circ \pm 3^\circ, 30^\circ \pm 3^\circ$) \citep{Kogut} in galactic coordinates. 
The CMB dipole, $(l,b)$ = ($263.99^\circ \pm 0.14^\circ, 48.26^\circ \pm 0.03^\circ$) from \citet{CMB_dipole_obs}, 
is caused by the motion of the Sun relative to the CMB, while the radio dipole, 
$(l,b)$ = ($248^\circ \pm 28^\circ, 46^\circ \pm 19^\circ$) from \citet{firstpaper},  
can be expected to receive contributions from the motion of the Solar System with respect to the CMB 
(kinetic dipole) and due to the uneven galaxy distribution (structure dipole). Within the current  accuracy, 
the direction of the radio dipole agrees with the CMB direction as well as with the motion of the 
Local Group with respect to the CMB. Therefore we expect the contribution of a local void to the radio dipole 
to add up with the velocity dipole, resulting in a larger dipole amplitude in radio surveys. 

The local structures considered in this work are not in conflict with the Copernican principle, as they are 
much smaller than the Hubble scale and thus a fine tuning of the position of the observer with respect to the centre of a void is not required. This is different to scenarios in which huge voids have been invoked to provide an 
alternative explanation of dark energy (e.g. \citet{Celerier}, \citet{Alnes05}, \citet{Alnes06}).

In this work, we will investigate this chain of thought in a more quantitative manner. Our model will 
be discussed in section \ref{Model}, followed by detailed testing in section \ref{Testing}. In section 
\ref{MissingDipole} we will examine the effects of realistic voids on the dipole, and we will present 
our conclusions in section \ref{Conclusion}.

\section{Model}
\label{Model}

For simplicity we model the observed universe limited by a radius of $R_{\rm o}$
 and with constant mean number density of sources everywhere (except in the area occupied by the void).  Therefore the results of this section can not directly be compared to radio surveys. The more realistic case 
 of a flux limited observation, with certain number counts, is discussed in section \ref{MissingDipole}. 

\begin{figure}[!t]
\begin{center}
 \includegraphics[width=6cm]{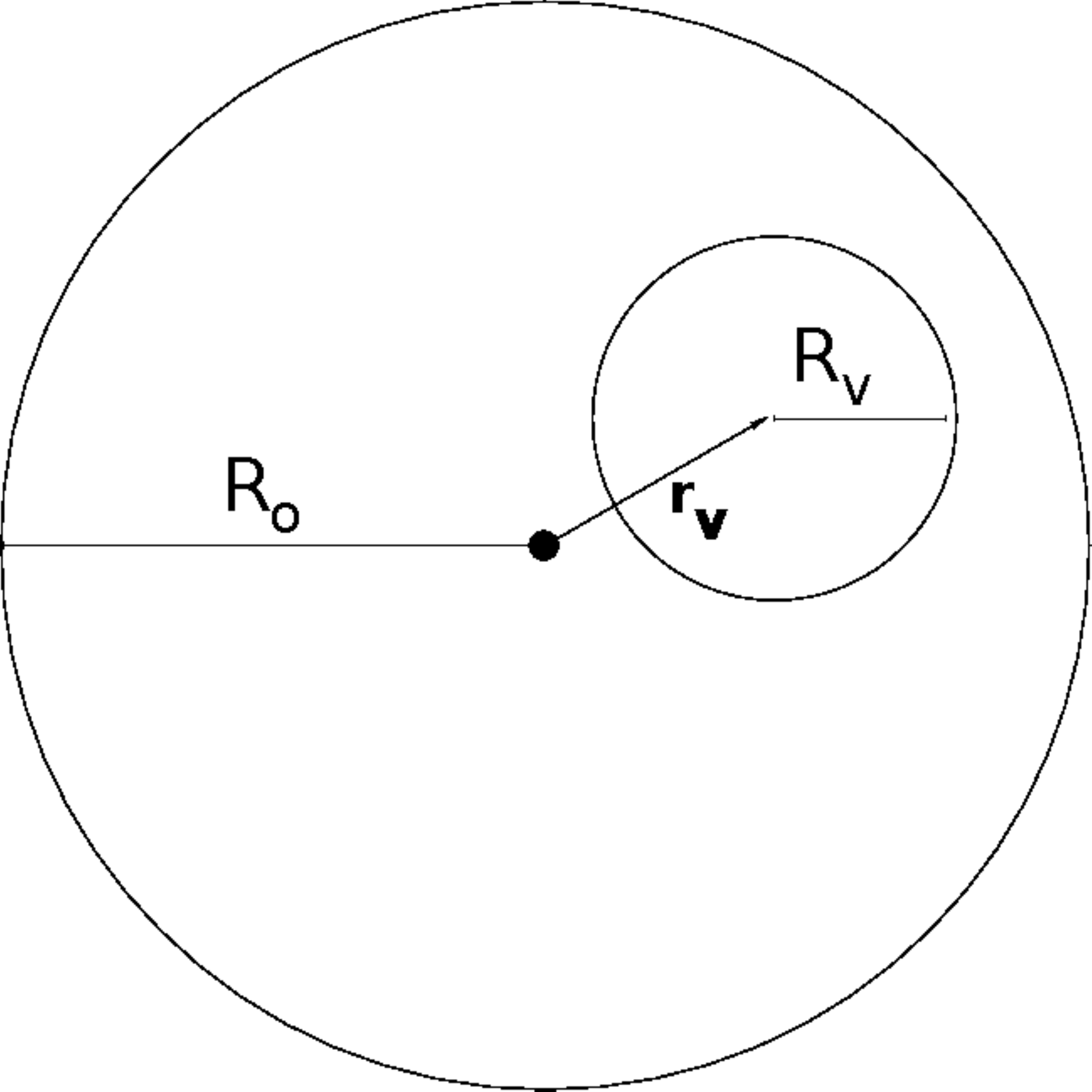} 
\caption{Configuration of our model of the observed volume-limited universe (radius $R_{\rm o}$) with a void of 
size $R_{\rm v}$ at distance $r_{\rm v}$ from the observer. \label{Setup}}
\end{center}
\end{figure}

The configuration of our model can be seen in figure \ref{Setup}. We consider a density contrast 
$\delta(r)$ in a region with radius $R_{\rm v}$, which we will call a void 
(but could be any amount of over- or under-density). We can restrict the calculation to 
the regions where $\delta(r) \neq 0$, as the contribution of the mean density to the dipole amplitude 
vanishes due to isotropy.

For the dipole measurement we use the linear estimator introduced by \cite{Crawford},
\begin{equation}
\label{Crawford}
 \vec d = \frac{1}{N} \sum_{i=1}^{N} \vec{\hat r}_i ,
\end{equation}
where $\vec{\hat r}_i$ is the normalized direction of source $i$ on the sky as seen by an observer in the 
centre of the observed universe. The fact that this estimator is linear is a big advantage here, since we can sum 
up the contributions of the background, of voids and of over-densities in an additive way. With a quadratic 
estimator this would not work out so trivially.

In order to simplify the integration, we pick a coordinate system centered on the void. The expectation of 
the observed dipole from the void, measured with the estimator (\ref{Crawford}), will be 
\begin{equation} 
\label{generaldipoleexp}
\langle \vec d \rangle = \bar{\alpha} \int_0^{2 \pi} \!\! {\rm d}\varphi \int_{-1}^1\!\! {\rm d} \cos{\vartheta}
 \int_0^{R_{\rm v}}\! {\rm d} r \, \delta(r)r^2  \frac{\vec{r}-\vec{r}_{\rm v}}{|\vec{r}-\vec{r}_{\rm v}|} \ .
\end{equation}
Here we have a normalization factor $\bar{\alpha}$. 

As a first case, we assume a constant density contrast $\delta$ in the void, and an offset 
$\vec{r}_{\rm v}$ of the void in direction $\vec{\hat{z}}$,
\begin{equation} 
\langle d_z \rangle = 
\bar{\alpha}\,  \delta \int_0^{2 \pi}\!\! {\rm d}\varphi \int_{-1}^1\!\! {\rm d} \cos{\vartheta} 
\int_0^{R_{\rm v}}\!\! {\rm d}r\,   r^2   
\frac{r \cos{\vartheta} -r_{\rm v}}{\sqrt{r^2 - 2 \cos{\vartheta} r r_{\rm v} +r_{\rm v}^2}}.
\end{equation}
This leads to
\begin{eqnarray} 
\label{OneVoid}
\nonumber \langle \vec{d} \rangle = \frac{4 \pi}{3} \bar{\alpha}\, \vec{\hat{r}}_{\rm v}\, \delta\,  R^3_{\rm v} 
[ \Theta(R_{\rm v} - r_{\rm v}) ( \frac{r_{\rm v}}{R_{\rm v}}- \frac{1}{5} \frac{r_{\rm v}^3}{R_{\rm v}^3} ) \\ 
+ \Theta(r_{\rm v} - R_{\rm v}) (1- \frac{1}{5} \frac{R_{\rm v}^2}{r_{\rm v}^2} ) ] \ ,
\end{eqnarray}
where $\Theta$ is the Heaviside function. This formula provides the dipole contribution of a top hat over- or underdensity for an observer inside or outside the void. 

Our aim is to investigate void regions with arbitrary density contrast profiles $\delta(r)$. In order to do so, we can heuristically linearly add up a large number $N$ of these voids to get to a smooth distribution $\delta (r)$. 

The normalization factor $\bar{\alpha}$ in (\ref{OneVoid}) can be found by the requirement that the integration over a sphere (with radius $R_{\rm o}$ bigger than the void size $R_{\rm v}$) should equal unity,
\begin{equation}
 1 = \bar{\alpha} 4 \pi \left( \int_{R_{\rm v}}^{R_{\rm o}} \mathrm{d}r\, r^2 + \int_0^{R_{\rm v}} \mathrm{d}r\, 
 r^2 (1+\delta(r))\right), 
\end{equation}
leading to
\begin{equation}
\label{alpha}
 \bar{\alpha} = \frac{3}{4\pi} \frac{1}{R_{\rm o}^3+3 \int_0^{R_{\rm v}} \mathrm{d}r\, r^2 \delta(r)} \ .
\end{equation}
We can see that the prefactor $\frac{3}{4 \pi}$ cancels in (\ref{OneVoid}). For convenience we 
introduce $\alpha=\frac{4 \pi}{3} \bar{\alpha}$ for all following formulae.

Let us consider the limit of a distant void $r_{\rm v} \gg R_{\rm v}$. Then we obtain
\begin{equation}
 \lim_{r_{\rm v} \gg R_{\rm v}}\langle d_z \rangle = \delta  \frac{R^3_{\rm v}}{R_{\rm o}^3} \ .
\end{equation}
So the dipole amplitude due to a void depends on the density contrast of the void 
and on the fraction of volume it occupies in the observed universe. 

For a realistic case of a flux limited observation of the universe, we need to generalise this formula to
\begin{equation}
 \lim_{r_{\rm v} \gg R_{\rm v}}\langle d_z \rangle \approx \delta  \frac{\tilde{N}_{\rm v}}{N_{\rm o}} \ .
\end{equation}
Here $N_{\rm o}$ is the number of sources in the observed universe and $\tilde{N}_{\rm v}$ is the number of sources we
 expect to see in the area occupied by the void, if it had the same mean number density as the
 rest of the universe. This number does depend on the flux limit, on the functional shape of the number counts 
and on the distance and size of the void.

\subsection{Observers outside the void}
\label{Outside of the Void}

Now we want to derive the expectation value of the dipole amplitude from voids with a density contrast 
$\delta(r)$, which is not constant. To do so, we will add up $N$ concentric voids, resulting in a
structure of $N$ concentric shells, each with constant density contrast $\delta_i, i = 1, \dots, N$.
The shells are ordered by their radius, starting at the shell with the biggest radius (shell number 1). 

We only look at the absolute value of $\langle \vec d \rangle$, since for symmetry reasons, the direction of 
this expectation value will always be $\vec{\hat r}_{\rm v}$. First we look at $N$ voids as observed from 
outside the voids, thus $r_{\rm v}>R_{\rm v}$. The second term in (\ref{OneVoid}) will give $N$ terms, 
which can be written as 
\begin{eqnarray}
\nonumber | d_z | &=&  \alpha \left[ \delta_1 R_1^3 (1 - \frac{1}{5} \frac{R_{1}^2}{r_{\rm v}^2})+ 
(\delta_2-\delta_1) R_2^3 (1 - \frac{1}{5} \frac{R_{2}^2}{r_{\rm v}^2}) + \right. \\  
&  & \left. ... + (\delta_N-\delta_{N-1}) R_N^3 (1 -\frac{1}{5} \frac{R_{N}^3}{r_{\rm v}^2}) \right] \ .
\end{eqnarray}
From this we obtain
\begin{eqnarray}
\nonumber |d_z| &=& 
\alpha \delta_N R_N^3 (1 - \frac{1}{5} \frac{{R_{N}^2}}{r_{\rm v}^2}) + \\ 
& & \alpha \sum_{i=1}^{N-1} \delta_i (R_{i}^3 - \frac{1}{5} \frac{R_{i}^5}{r_{\rm v}^2}-R_{i+1}^3 + \frac{1}{5} \frac{R_{i+1}^5}{r_{\rm v}^2}) \ .
\end{eqnarray}
Now we take the difference in size between consecutive shells to be infinitesimally small, 
meaning $R_{i+1}=R_{i} - \epsilon$. Without loss of generality, we can put $R_{1}=r_{\rm v}$ and therefore
place the observer on the edge of the biggest void shell 
(if $r_{\rm v} > R_{\rm v}$ then $\delta(r)|_{r>R_{\rm v}}=0$). The innermost void shell will have a 
vanishing radius and so $R_{N}=0$. This leads to
\begin{equation}
 |d_z|=\alpha  \sum_{i=1}^{N-1} \delta_i (3 R_{i}^2 \epsilon - \frac{R_{i}^4 \epsilon}{r_{\rm v}^2}) \ ,
\end{equation}
which can be written in the form of an integral
\begin{equation}
\label{outwardint}
 |d_z| = \alpha \int_0^{r_{\rm v}}\!\! \rm{d} r\,  r^2 \delta(r) \left(3 - \frac{r^2}{r_{\rm v}^2}\right).
\end{equation}
This is the equation we have been seeking for the dipole observed by an observer outside the void.

\subsection{Observers inside the void}
\label{Inside of the Void}

Now we examine the case of $N$ void shells, each of constant density, with $r_{\rm v} \leq R_{\rm v}$; the  observer is inside the void. We have
\begin{eqnarray}
 \nonumber | d_z | &=& \alpha \left[ \delta_1(R_{1}^2r_{\rm v}- \frac{1}{5} r_{\rm v}^3) + 
 (\delta_2-\delta_1)(R_{2}^2r_{\rm v}- \frac{1}{5} r_{\rm v}^3)+ \right. \\
 &&  \left.   ...+ (\delta_N-\delta_{N-1})(R_{N}^2r_{\rm v}- \frac{1}{5} r_{\rm v}^3) \right].
\end{eqnarray}
This can be rewritten as
\begin{equation}
 |d_z|=\alpha \delta_N (R_{N}^2r_{\rm v} - \frac{1}{5} r_{\rm v}^3) + 
 \alpha \sum_{i=1}^{N-1} \delta_i (R_{i}^2r_{\rm v} - R_{i+1}^2r_{\rm v}) \ .
\end{equation}
Again we make the difference in size between consecutive void shells infinitesimally small, 
meaning $R_{i+1}=R_{i}-\epsilon$. The shell with the smallest radius that still includes the 
observer ($r_{\rm v} \leq R_{\rm v}$), will have $R_{vN}=r_{\rm v}$ and $\delta_N=\delta(r_{\rm v})$. The void shell with the biggest radius will have $R_{1}=R_{\rm v}$. This leads to
\begin{equation}
 |d_z|= \frac{4}{5} \alpha \delta(r_{\rm v})  r_{\rm v}^3 - 2 \alpha r_{\rm v} \sum_{i=1}^{N-1} \delta_i R_{vi} \epsilon,
\end{equation}
which can be written as an integral
\begin{equation}
\label{inwardint}
 |d_z|= \frac{4}{5} \alpha  r_{\rm v}^3  \delta(r_{\rm v}) + 
 2 \alpha r_{\rm v} \int_{r_{\rm v}}^{R_{\rm v}}\!\! \mathrm{d}r\,  r \delta(r).
\end{equation}
This is the form we have been seeking for the dipole observed when the observer is inside a void.

\subsection{Structure dipole amplitude}

When combining the results for an observer inside a void (section \ref{Inside of the Void}) and those for an 
observer outside the void (section \ref{Outside of the Void}), we need to be careful. The void shell at the 
position of the observer $r_{\rm v}$ and density contrast $\delta(r_{\rm v})$ has been counted in 
both cases. The formula for an observer outside the void (\ref{outwardint})
gives $\frac{4}{5}\alpha \delta(r_{\rm v}) r_{\rm v}^3$, which is the same result we find for the 
observer inside the void (\ref{inwardint}) with $\delta(r)|_{r<R_{\rm v}}=\delta(r_{\rm v})$. Therefore we need 
to subtract this term once when combining both cases. We obtain
\begin{eqnarray}
\label{intform}
\nonumber \langle \vec d \rangle &=& 
\alpha \vec{\hat r}_{\rm v} \int_0^{\mathrm{min}(r_{\rm v},R_{\rm v})}\!\! \rm{d} r\,  r^2 \delta(r) 
(3-\frac{r^2}{r_{\rm v}^2}) + \\ && 
\alpha \vec{\hat r_{\rm v}}  \Theta(R_{\rm v}-r_{\rm v})\,  2  r_{\rm v} \int_{r_{\rm v}}^{R_{\rm v}}\!\! 
\mathrm{d}r\, r \delta(r).
\end{eqnarray}
The upper boundary of the first integral is now the minimum of $r_{\rm v}$ and $R_{\rm v}$, since in general it is not guaranteed that $r_{\rm v}<R_{\rm v}$.

\section{Testing}
\label{Testing}

We test our mathematical model with the help of computer simulations. The focus of the first subsection below is 
to verify the dependence of a dipole contribution on the three void parameters $R_{\rm v}$, $r_{\rm v}$ and $
\delta$. Next we allow for a varying density contrast $\delta$ with respect to $r$. Up to this point, we assume a 
volume limited observation. The flux limited case, including realistic number counts, is discussed in 
section \ref{MissingDipole}, where we incorporate a radio sky simulation from \citet{SSS}.

\subsection{Structures of constant density contrast}
\label{cdc}

Let us first look at constant density contrasts $\delta(r)=\delta$ inside the void area. In order to test our
calculations, we construct a simple simulation. We draw a random point (with the random number 
generator Mersenne Twister) inside a three dimensional sphere of radius $R_{\rm o}$, which we set to 
$R_{\rm o}=1$ (which fixes the physical scale). The points inside this sphere are 
uniformly distributed. 

The next step depends on whether we have an underdensity ($\delta<0$) or an overdensity ($\delta>0$) 
of radius $R_{\rm v}$. In the first case, we keep all points which are outside the void 
(this represents the average density of objects, i.e. $\delta=0$). For each point inside the void,  
we draw a random number between $0$ and $1$. If this number is bigger than $\delta+1$ we drop this 
point and turn to the next one. If, on the other hand, it is smaller than $\delta +1$, we keep it and proceed 
to a new point (this algorithm is simply a Monte Carlo sampling between $\delta=-1$ and $\delta=0$). 

For the case $\delta>0$, we keep all drawn points inside the overdensity, and draw random numbers 
($0 \rightarrow 1$) for points outside the overdensity. Now we drop the point only if the random number is 
larger than $1/(1+\delta)$. So we create a map with the desired densities inside and outside the 
over-/underdense region. 

In this way we will draw $N$ points in total, which will be used to measure $\vec d$ via (\ref{Crawford}). 
Due to the fact that we can only use finite values of $N$, our simulation will always have a certain amount 
of shot noise, whereas our calculations in Section 2 neglected noise. In \cite{firstpaper} the influence of this 
shot noise on the expectation value of a linear estimator is discussed. We compare the average outcome of 
several simulations with
\begin{equation}
\label{shotnoise}
 \tilde d := \sqrt{\langle \vec{d} \rangle^2 + (0.92/\sqrt{N})^2} \ ,
\end{equation}
where the second term inside the square root comes from the shot noise contribution. For 
$\langle \vec d \rangle$ we can use the results discussed in section 2, 
depending on the case we are simulating.

\begin{table}
\begin{tabular}{cccccc}
 $r_{\rm v}$ &  $R_{\rm v}$ &  $\delta$  & $\tilde d$ & $d_s$ & error\\
  & &   & ($10^{-2}$)  & ($10^{-2}$) & $\%$ \\ \hline \hline
0.1 & 0.1 & -1 & 0.12 & 0.13 & 8,7 \\ \hline
0.1 & 0.2 & -1 & 0.39 & 0.41 & 2,8 \\ \hline
0.2 & 0.3 & -1 & 1.69 & 1.71 & 1,3 \\ \hline
0.2 & 0.4 & -1 & 3.25 & 3.28 & 0,9 \\ \hline
0.4 & 0.4 & -1 & 5.47 & 5.47 & 0,0 \\ \hline
0.1 & 0.1 & -0.5 & 0.10 & 0.11 & 10,1 \\ \hline
0.1 & 0.2 & -0.5 & 0.21 & 0.17 & 20,2 \\ \hline
0.2 & 0.3 & -0.5 & 0.84 & 0.84 & 0,2 \\ \hline
0.2 & 0.4 & -0.5 & 1.57 & 1.54 & 2,1 \\ \hline
0.4 & 0.4 & -0.5 & 2.65 & 2.68 & 1,3 \\ \hline
0.1 & 0.1 & 1 & 0.12 & 0.13 & 5,7 \\ \hline
0.1 & 0.2 & 2 & 0.75 & 0.76 & 1,4 \\ \hline
0.2 & 0.3 & 4 & 5.92 & 5.89 & 0,5 \\ \hline
0.2 & 0.4 & 5 & 11.52 & 11.50 & 0,1 \\ \hline
0.4 & 0.4 & 7 & 24.75 & 24.70 & 0,2 \\ \hline
\hline
\end{tabular}
\caption{Comparison of analytic model and simulation for an observer inside a local 
spherical structure ($r_{\rm v} < R_{\rm v}$) with constant density contrast $\delta$. 
The analytically calculated dipole is denoted by $\tilde d$. Each simulated dipole amplitude $d_s$ is an average 
of 10 simulations with $N=10^6$ sources each; the error is defined as $2|(\tilde d - d_s)/(\tilde d + d_s)|$.}
\label{inwardtophats}
\end{table}

\begin{table}
\begin{tabular}{cccccc}
 $r_{\rm v}$ &  $R_{\rm v}$ &  $\delta$  & $\tilde d$ & $d_s$ & error\\
  & &   & ($10^{-2}$)  & ($10^{-2}$) & $\%$ \\ \hline \hline
0.1 & 0.1 & -1 & 0.12 & 0.10 & 18.7 \\ \hline
0.2 & 0.1 & -1 & 0.13 & 0.13 & 3.3 \\ \hline
0.3 & 0.2 & -1 & 0.74 & 0.73 & 1.7 \\ \hline
0.4 & 0.2 & -1 & 0.77 & 0.77 & 0.9 \\ \hline
0.4 & 0.4 & -1 & 5.47 & 5.46 & 0.2 \\ \hline
0.1 & 0.1 & -0.5 & 0.10 & 0.11 & 7.4 \\ \hline
0.2 & 0.1 & -0.5 & 0.10 & 0.09 & 12.9 \\ \hline
0.3 & 0.2 & -0.5 & 0.38 & 0.40 & 5.8 \\ \hline
0.4 & 0.2 & -0.5 & 0.39 & 0.36 & 8.4 \\ \hline
0.4 & 0.4 & -0.5 & 2.65 & 2.65 & 0.1 \\ \hline
0.1 & 0.1 & 1 & 0.12 & 0.12 & 3.2 \\ \hline
0.2 & 0.1 & 2 & 0.21 & 0.21 & 2.3 \\ \hline
0.3 & 0.2 & 4 & 2.83 & 2.81 & 0.6 \\ \hline
0.4 & 0.2 & 5 & 3.66 & 3.66 & 0.1 \\ \hline
0.4 & 0.4 & 7 & 24.75 & 24.70 & 0.2 \\ \hline
\hline
\end{tabular}
\caption{As table \ref{inwardtophats}, but for observers sitting outside the spherical structure.}
\label{outwardtophats}
\end{table}

In table (\ref{inwardtophats}) we see a comparison between our analytic expectation and the 
simulated results, for cases where the observer is inside the void. In order to quantify the performance of 
the theory we estimate the error by $2|(\tilde d - d_s)/(\tilde d + d_s)|$. We see in table (\ref{inwardtophats}) that 
this error drops as the dipole values increase. This is due to the fact that in those cases the uncertainties due to 
shot noise are less important. For the case of $r_{\rm v}=0.1  ,\ R_{\rm v}= 0.2 \ \rm{and} \ \delta = -0.5$ we see an 
unusually high error. We repeated this configuration with 20 extra simulations and found an averaged value of 
$d_s=0.215 \times 10^{-2}$, which is very close to $\tilde d$; so we are confident that this relatively large 
disagreement arose by chance. 
In all other cases we see a good agreement between the calculated values and the simulated ones. If the dipole 
is  large, the agreement becomes remarkably good. These results confirm the calculated expectation values 
of the dipole for voids with $r_{\rm v} \le R_{\rm v}$ and constant density contrast $\delta$.

In table (\ref{outwardtophats}) we present the comparison for cases with $r_{\rm v} \ge R_{\rm v}$. Again 
we can see that the difference between calculation and simulation is quite small, and decreases as the dipole 
amplitude increases.

\begin{figure}
 \includegraphics[width=6cm,angle=270]{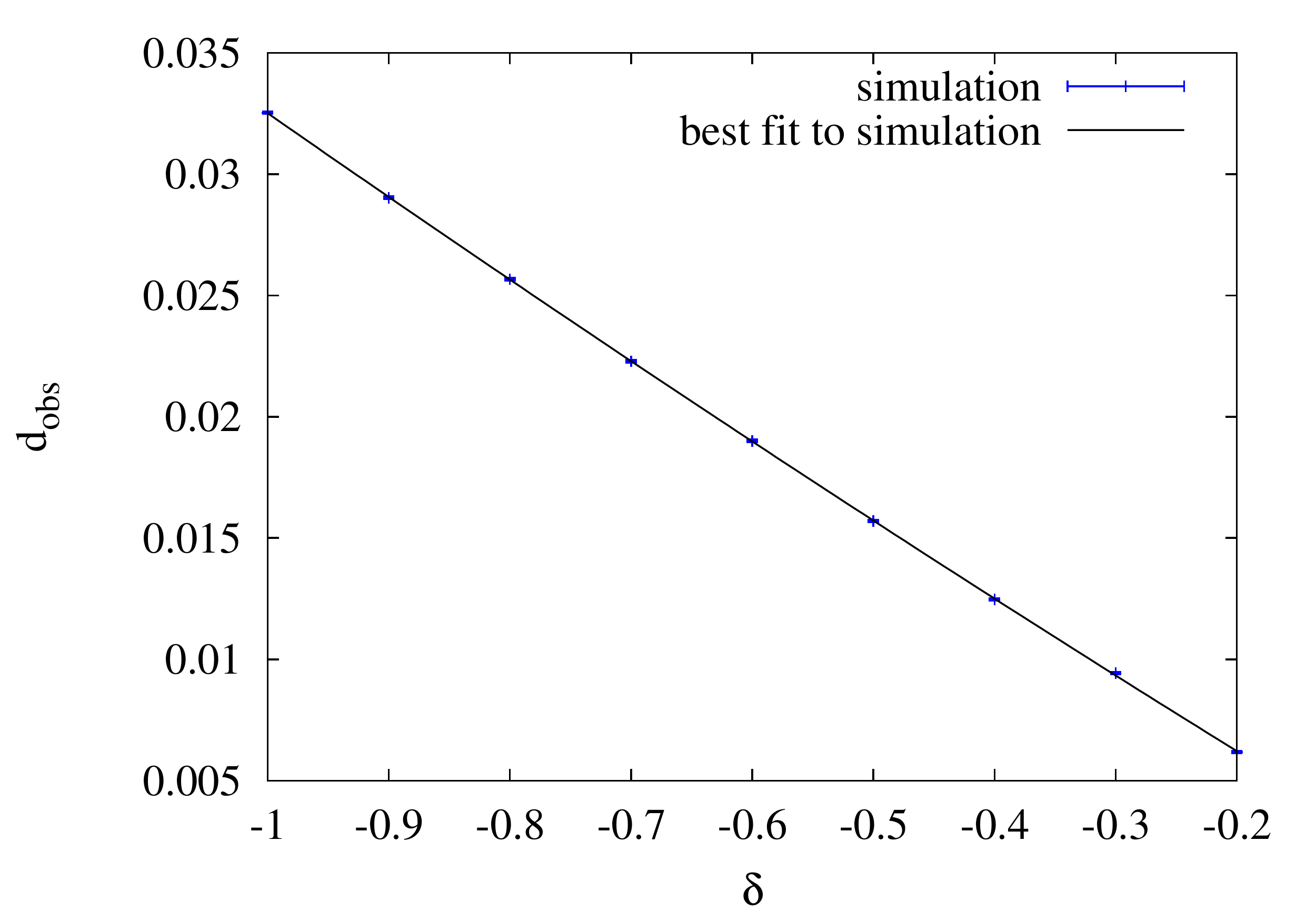}
\caption{Simulated dipole amplitudes. The graph is for a void with $R_{\rm v}=0.4$ and $r_{\rm v}=0.2$ for 
different values of $\delta$, while the curve is the best fit. Each data point is the mean value of the dipole 
amplitude from $100$ simulations with $10^6$ sources each. The error bars represent the empirical variance of 
these simulations. For the fit a function $f(\delta)=\sqrt{\left(\delta \frac{a}{1+b \delta} \right)^2+c^2}$ was used.}
\label{Graphdelta}
\end{figure}

\begin{figure}
 \includegraphics[width=6cm,angle=270]{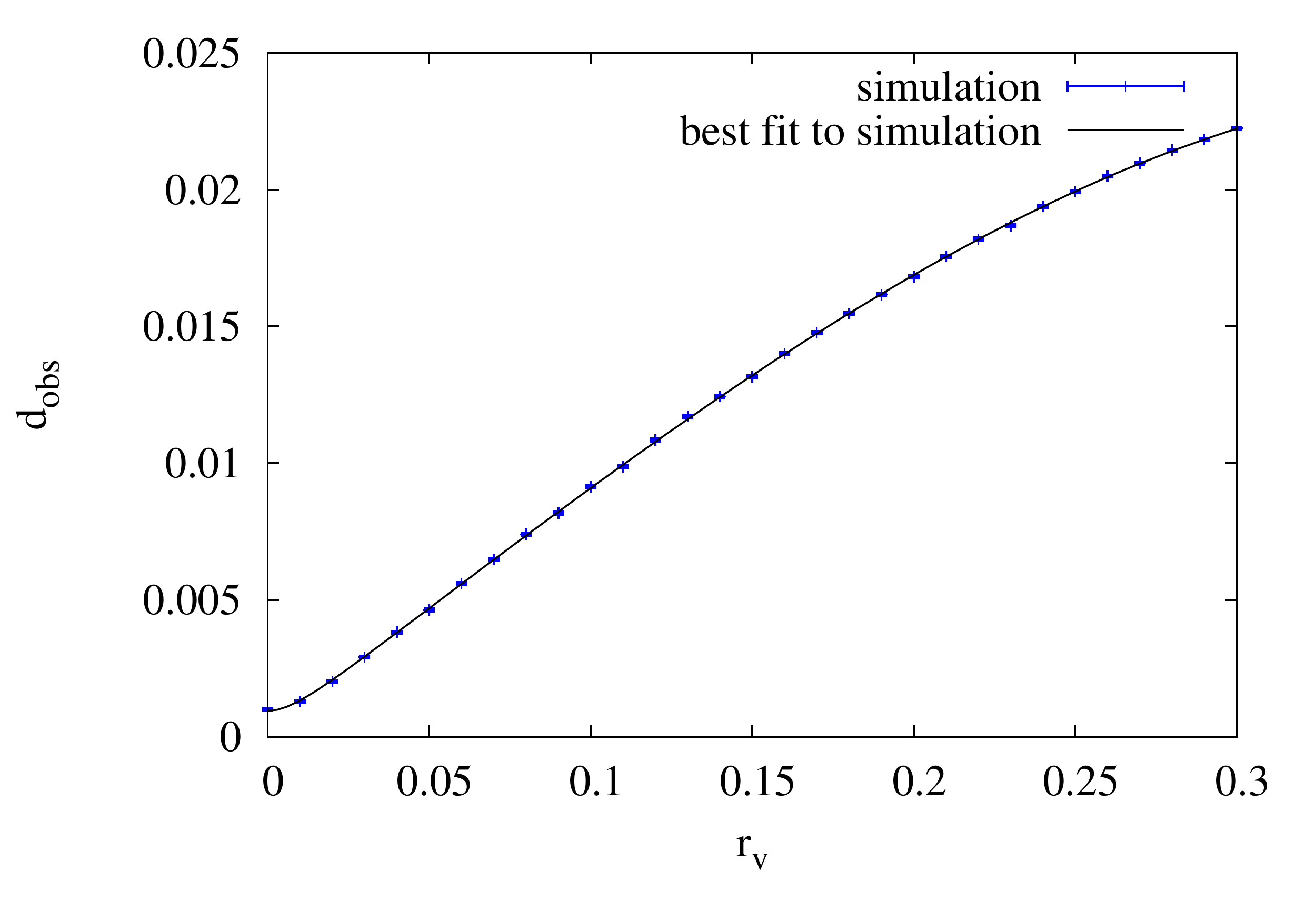}
\caption{Simulated dipole amplitudes. The graph is for a void with $R_{\rm v}=0.3$ and $\delta=-1$ for different 
values of $r_{\rm v}$. Each data point is the mean value of the dipole amplitude from $100$ simulations with 
$10^6$ sources each, while the curve is the best fit. The error bars represent the empirical variance of these 
simulations. For the fit, a function $g(r_{\rm v})=\sqrt{(a \ r_{\rm v} - b \ r_{\rm v}^3)^2+c^2}$ was used.
\label{Graphrv}}
\end{figure}

\begin{figure}
 \includegraphics[width=6cm,angle=270]{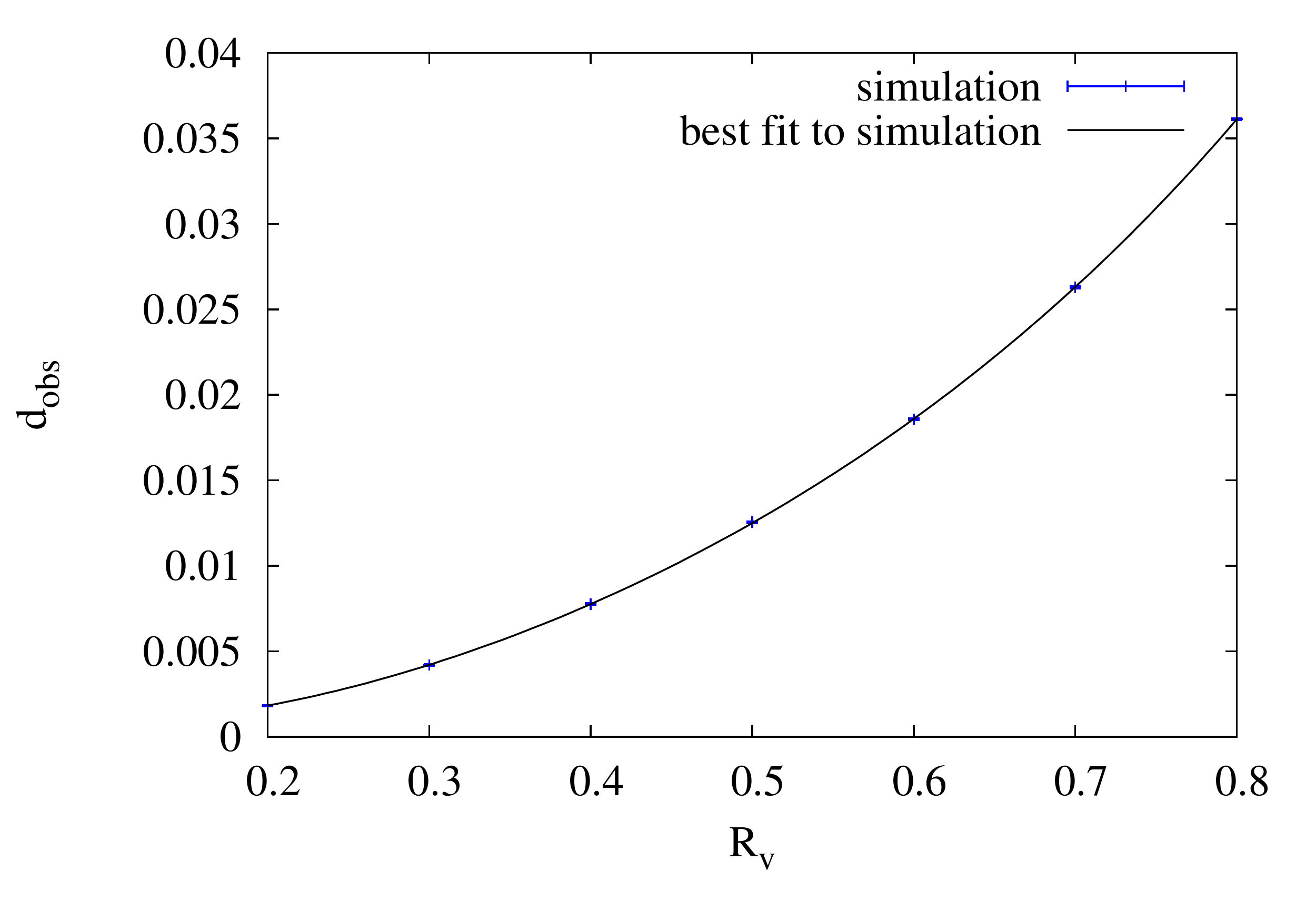}
\caption{Simulated dipole amplitudes. The graph is for a void with $r_{\rm v}=0.2$ and $\delta=-0.25$ for 
different values of $R_{\rm v}$. Each data point is the mean value of the dipole amplitude from $100$ 
simulations with $10^6$ sources each, while the curve is the best fit. The error bars represent the empirical 
variance of these simulations. For the fit, a function $h(R_{\rm v})=\sqrt{\left( b\frac{a\ R_{\rm v}^2-0.2a^3}{1+b \ 
R_{\rm v}^3} \right)^2+c^2}$ was used.\label{GraphRv}}
\end{figure}

The simulated dipole amplitude can be plotted as a function of either $r_{\rm v}$, $R_{\rm v}$ or $\delta$. We 
present examples of simulations in  figures \ref{Graphdelta}, \ref{Graphrv} and \ref{GraphRv}, 
where we have 
fitted functions of the form (\ref{OneVoid}) making use of the normalization factor (\ref{alpha}) and including a shot noise contribution (\ref{shotnoise}).

In all cases the fitted curve follows the simulated dipole amplitudes very well. The first case shows the 
dipole amplitude as a function of the density contrast $\delta$; we see that the dependence on $\delta$ is 
approximately linear. Here we used a void of size $R_{\rm v}=0.4$ and an offset distance of $r_{\rm v}=0.2$; 
we expect from our theoretical model fit parameters of $\langle a \rangle= 0.0305 $, 
$\langle b \rangle = 0.064$ and $\langle c \rangle = 0.92 \times 10^{-3}$. The values of our fit of 
$f(\delta)$ give us the parameters $a=0.0303 \pm 0.0002$, $b=0.066 \pm 0.005$ and 
$c=(0.92 \pm 0.05) \times 10^{-3}$, which are in excellent agreement. 

For figure \ref{Graphrv} we used $10^6$ sources, a density contrast of $\delta=-1$ and a void radius of 
$R_{\rm v}=0.3$; we expect $\langle a \rangle = 0.0925 $, $\langle b \rangle = 0.206$ and 
$\langle c \rangle = 0.92 \times 10^{-3}$. The values of our fit of $g(r_{\rm v})$ give us the parameters 
$a=0.0924 \pm 0.0002$, $b=0.205 \pm 0.003$ and $c= (0.92 \pm 0.04) \times 10^{-3}$. 
Again, this is in very good agreement with our prediction. We can observe that the dipole increases 
strongly with the offset distance $r_{\rm v}$. On the edge of the void, the increase becomes more modest.

The graph in figure \ref{GraphRv} shows the behaviour of the dipole amplitude as a function of the void size 
$R_{\rm v}$. Here we used a density contrast of $\delta=-0.25$ and an offset distance of $r_{\rm v}=0.2$; 
we expect $\langle a \rangle= 0.2 $, $\langle b \rangle = 0.25$ and $\langle c \rangle = 0.92 \times 10^{-3}$. 
The values of our fit of $h(R_{\rm v})$ give us the parameters $a=0.205 \pm 0.006$, $b=0.244 \pm 0.006$ 
and $c= (0.91 \pm 0.07) \times 10^{-3}$. We see that the parameters are in very good agreement with 
our prediction.

We conclude from this section that formula ($\ref{OneVoid}$) combined with (\ref{shotnoise}) is in very good 
agreement with our simulations.

\subsection{Arbitrary void profile}

Now we would like to test whether formula (\ref{intform}) is also verified by our simulations. We consider 
density contrasts of the form $\delta(r)=\frac{r^p}{R_{\rm v}^p}-1$ for $r$ inside the void and 
$\delta(r)=0$ outside. In all such cases the density contrast has the boundary values $\delta(0)=-1$ and 
$\delta(R_{\rm v})=0$. In such cases the integrals in (\ref{intform}) can be solved analytically. 
Results will be put into (\ref{shotnoise}) in order to get $\tilde d$. 

The simulation is similar to the one described in section \ref{cdc}. We choose a random number for points 
inside the void. This time a point (with distance $r$ from the void centre)  is discarded if the random number 
is greater than $r^p/R_{\rm v}^p$.

\begin{table}
\begin{tabular}{cccccc}
 $r_{\rm v}$ &  $R_{\rm v}$ &  $p$  & $\tilde d$ & $d_s$ & error\\
  & &   & ($10^{-2}$)  & ($10^{-2}$) & $\%$ \\ \hline \hline
0.2 & 0.4 & 1 & 0.96 & 0.96 & 0.2 \\ \hline
0.3 & 0.5 & 1 & 2.16 & 2.18 & 0.7 \\ \hline
0.2 & 0.4 & 2 & 1.49 & 1.51 & 1.4 \\ \hline
0.3 & 0.5 & 2 & 3.42 & 3.44 & 0.4 \\ \hline
0.2 & 0.4 & 3 & 1.82 & 1.83 & 0.2 \\ \hline
0.3 & 0.5 & 3 & 4.24 & 4.25 & 0.2 \\ \hline
\hline
\end{tabular}
\caption{Comparison for cases with offset distance $r_{\rm v}$ smaller than void radius $R_{\rm v}$. The density 
contrast is of the form $\delta(r)=\frac{r^p}{R_{\rm v}^p}-1$ and the calculated dipole is $\tilde d$. Each simulated 
dipole amplitude $d_s$ is an average of 10 simulations with $N=10^6$ sources each; the error is defined as $2|
\frac{\tilde d - d_s}{\tilde d + d_s}|$.}\label{inwardforms}
\end{table}

\begin{table}
\begin{tabular}{cccccc}
 $r_{\rm v}$ &  $R_{\rm v}$ &  $p$  & $\tilde d$ & $d_s$ & error\\
  & &   & ($10^{-2}$)  & ($10^{-2}$) & $\%$ \\ \hline \hline
0.4 & 0.2 & 1 & 0.21 & 0.21 & 1.9 \\ \hline
0.5 & 0.3 & 1 & 0.65 & 0.69 & 4.7 \\ \hline
0.4 & 0.2 & 2 & 0.32 & 0.31 & 2.8 \\ \hline
0.5 & 0.3 & 2 & 1.04 & 1.05 & 1.0 \\ \hline
0.4 & 0.2 & 3 & 0.40 & 0.42 & 5.1 \\ \hline
0.5 & 0.3 & 3 & 1.30 & 1.30 & 0.0 \\ \hline
\hline
\end{tabular}
\caption{Comparison for cases with offset distance $r_{\rm v}$ larger than void radius $R_{\rm v}$. The density 
contrast is of the form $\delta(r)=\frac{r^p}{R_{\rm v}^p}-1$ and the calculated dipole is $\tilde d$. Each simulated 
dipole amplitude $d_s$ is an average of 10 simulations with $N=10^6$ sources each; the error is defined as $2|
\frac{\tilde d - d_s}{\tilde d + d_s}|$.}\label{outwardforms}
\end{table}

In table (\ref{inwardforms}) we see the comparison of our calculated dipole expectation with the simulation 
results  for cases $r_{\rm v}<R_{\rm v}$, and in table (\ref{outwardforms}) for cases with $R_{\rm v}>r_{\rm v}$. 
Again we can observe the tendency to find improved agreement with the analytic model
when the dipole amplitude is larger. 
In fact even for small dipole values we  see a good agreement between the simulation and our calculation. 
Therefore we are satisfied that equation (\ref{intform}) is confirmed by our simulations.

\section{Missing dipole contribution}
\label{MissingDipole}

Now we investigate the contribution which realistic void models can have on the observed radio dipole.
 Therefore we no longer assume a volume limited observation, but a flux limited one. 

In \cite{firstpaper}, a dipole amplitude $d_{\rm{radio}}=(1.8 \pm 0.6) \times 10^{-2}$ in the NVSS catalogue 
was reported, which is significantly above the prediction inferred from CMB measurements 
\citep{CMB_dipole_obs} of $d_{\rm{cmb}}= (0.48 \pm 0.04) \times 10^{-2}$. Therefore we can infer a 
missing dipole contribution 
\begin{equation}
\mathbf{\Delta \tilde{d} = \tilde{d}_{\rm{radio}} - d_{\rm{cmb}} =} (1.3 \pm 0.6) \times 10^{-2},
\end{equation}
where the tilde indicates that the dipole amplitudes include correction factors, 
which are also be applied to the void dipole estimations below.

We would like to investigate whether it is possible to get a dipole contribution of this magnitude from a void 
model in which we are off-centre. We examine a void of the type described by \cite{Void1}. They report an 
observed void with a density contrast of about $-\frac{1}{3}$ up to redshifts of about $z=0.07$. 
The influence of smaller voids, compared to \citet{Void1}, on the clustering dipole, was discussed e.g. by \citet{Bilicki}.

Since we want to compare the void dipole $d_{\rm{void}}$ with the one derived from the NVSS catalogue, we 
cannot assume a constant density outside the void. The NVSS itself does not contain information about the 
distance of individual objects. In order to have a realistic redshift distribution, we used the semi-empirical $S^3$ 
simulation from \cite{SSS} with an area of $400$ square degrees.
From this we obtained a catalogue of 
approximately $2800$ radio sources with flux densities above $25$ mJy at $1.4$ GHz (the 
limit \cite{firstpaper} used for obtaining $d_{\rm radio}$). This is now a flux limited 
observation, in contrast to the volume limited model in the previous section. 

Now we modify our void simulation in the following way. Each data point will have a randomly chosen 
direction on the sky and a redshift distance chosen from the $S^3$ catalogue. Inside the void we will 
reduce the density of points in the way described in section \ref{cdc}. So we are left with 
number counts outside the void, which are close to what is actually observed in the mean, and 
some density contrast $\delta(r)$ inside the void.

We would like to estimate the maximal contribution of such a void to the measured dipole amplitude. Therefore 
we choose $r_{\rm v}=R_{\rm v}$, since this will give the biggest dipole amplitude for a void which includes the 
observer. As we consider $r_{\rm v}$ to be much less than  the Hubble distance $R_{\rm H}$, we can use 
the linear Hubble law to relate distances and redshift. The shot noise in this simulation should be 
suppressed, since 
we only want to know the pure contribution from the void (any possible shot noise is already taken into 
account by the error bars of $d_{\rm{radio}}$). So we choose the following parameters for our simulation, which 
uses the redshift information to infer the distance parameters: $R_{\rm v}=0.07 R_{\rm H}$, $\delta = - 1/3$ 
and $N=10^7$. 
We carried out $50$ runs of our simulation, and the average dipole we obtained is
\begin{equation}
 d_{\rm{void}}= (0.0918 \pm 0.0023) \times 10^{-2}.
\end{equation}
For $r_{\rm v}=0.06 R_{\rm H} < R_{\rm v}$ we obtained $d_{\rm{void}}= (0.0839 \pm 0.0026) \times 10^{-2}$. Lower values of 
$r_{\rm v}$ lead to lower dipole amplitudes.

Due to masking effects (incomplete sky coverage and galactic foreground) the dipole amplitude 
in \cite{firstpaper} was multiplied by $3/k$, where $k$ was 
evaluated to be $1.34$ for this case. In order to compare both dipole amplitudes, we also need to multiply $d_
{\rm{void}}$ by this number,
\begin{equation}
 \tilde d_{\rm void}=\frac{3}{k} d_{\rm void} = (0.21 \pm 0.01) \times 10^{-2}.
\end{equation}

If we compare this $\tilde d_{\rm void}$ to the missing dipole $\Delta d$ we can see that such a void 
can have a significant contribution to the observed radio dipole. When we consider the lower bound of
$d_{\rm{radio}}$ we see that $\tilde d_{\rm{void}}$ could explain up to about a third of 
the missing radio dipole.

If we would like to explain the missing dipole contribution only by one void, this would need to be bigger and 
have a larger density contrast. One possible combination of void parameters would be 
$R_{\rm v} = r_{\rm v} = 0.11 R_{\rm H}$ and $\delta = - 0.6$. For this we found, with $50$ simulations, an 
average dipole (including masking corrections, described above) of
\begin{equation}
 \tilde d_{\rm{void}}=\frac{3}{k} d_{\rm{void}} = (0.72 \pm 0.01) \times 10^{-2}.
\end{equation}

A similar result, $d = (0.69 \pm 0.01) \times 10^{-2}$, is obtained with the parameters 
$R_{\rm v}= r_{\rm v}=0.15 R_{\rm H}$ and $\delta = - 1/3$. These void models are not particularly extreme. 
Another possibility would be to have a combination of different under- and over-densities, such as e.g. 
superclusters. The dipole of these different structures could add up to result in an amplitude which could 
potentially explain the whole missing dipole contribution.

\subsection{Flux and frequency dependence}

\begin{figure}
 \includegraphics[width=6cm,angle=270]{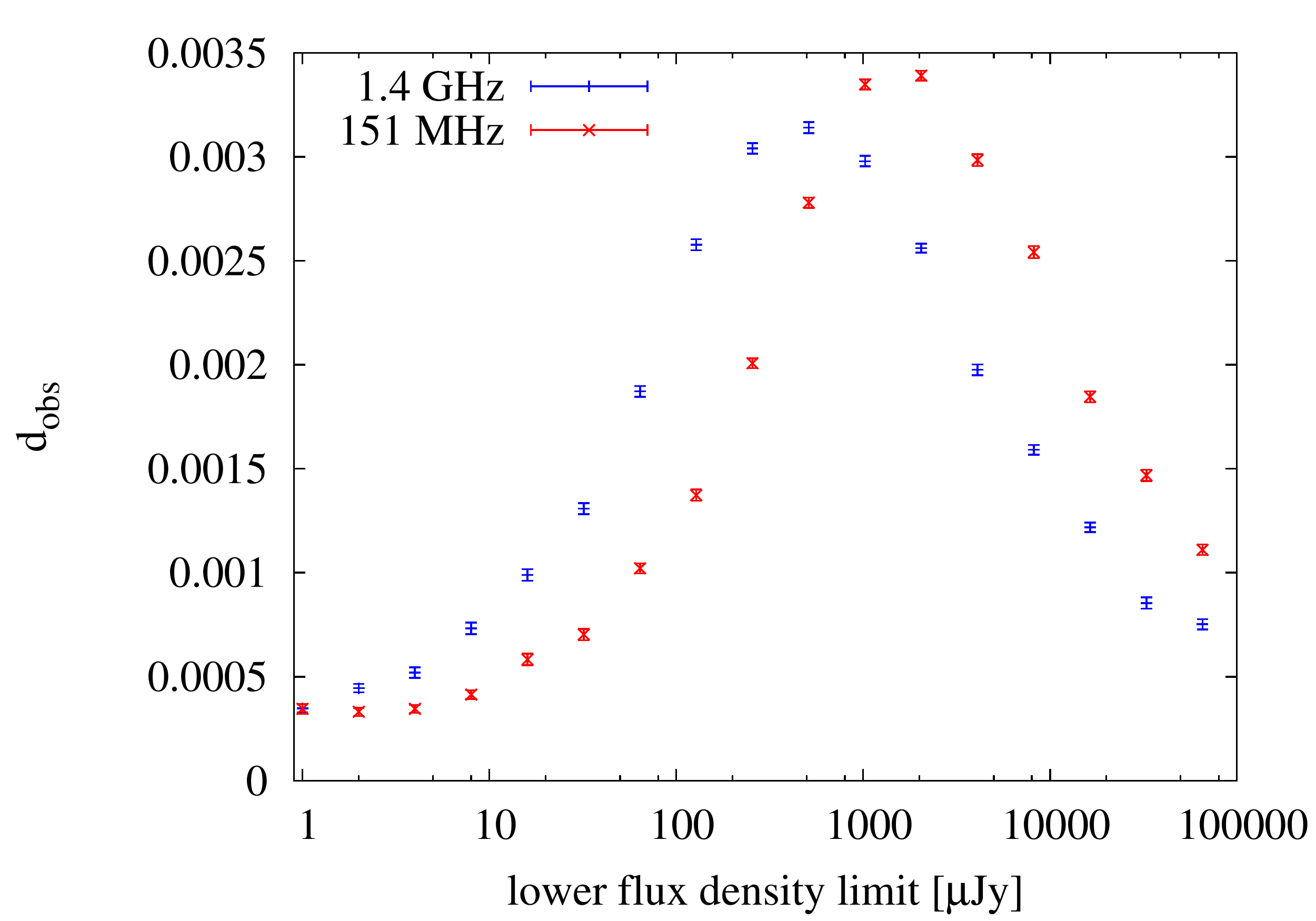}
\caption{Simulated dipole amplitudes for different flux limits. Each point is the average of $50$ simulations with 
$10^7$ sources each. The void used here has the parameters $R_{\rm v}= r_{\rm v}=0.07 R_{\rm H}$ and $\delta=-1/3$. 
The error bars  represent the empirical variance of these simulations.\label{GraphFlux}}
\end{figure}

So far we only used the $S^3$ simulation with a lower flux density limit of $25$ mJy. For future radio surveys we 
hope to be able to estimate the radio dipole with more sources and therefore we will need to apply a lower flux 
density limit. The effect of a change in this flux density limit for the dipole contribution of a void is not trivially 
estimated. On the one hand, a lower flux density limit means that we can see more distant sources then 
previously. This means that local structure becomes less important. On the other hand, a lower flux density limit 
will also lead to the detection of nearby galaxies, which have a low radio brightness. For those galaxies the void 
structure is important and we could expect an increase in the measured dipole amplitude. Both effects will vary in 
strength at different flux density ranges.

In order to estimate these effects, we again used the $S^3$ simulation from \cite{SSS}. We considered the 
two frequencies of $1.4$ GHz (e.g.~NVSS or a planned survey with 
ASKAP\footnote{www.atnf.csiro.au/projects/askap/}) and $151$ MHz 
(e.g.~LOFAR\footnote{www.lofar.org}). A continuum survey with 
the SKA\footnote{www.skatelescope.org} will be likely to be collected at a frequency between these 
(e.g. 600 MHz). Again we used the void parameters from \cite{Void1}. This time we applied different flux 
density limits to see the dependence of the observed dipole amplitudes on the flux density limit.

In figure \ref{GraphFlux} we see that the dependence of the measured dipole amplitudes from the flux density 
limit is quite complex. Notice that the flux limits shown cover almost five orders of magnitude.  
For flux density limits below $10$ mJy, the dipole amplitudes increase very strongly until a maximum is reached 
around $1$ mJy. The contribution of a local void for the dipole in a survey with a flux 
density limit of $1$ mJy could be about three times as strong as it is for a limit of $25$ mJy. This means that the effect 
of voids will become more important in future radio surveys. In principle it is possible to disentangle the kinetic 
dipole contribution from the structure dipole, since the kinetic dipole amplitude does not depend on the 
flux density limit (the shot noise does, but this will be taken into account by the error bars).

We can see that the general behaviour for both frequencies is the same. The main difference is in the position 
of the peak in the dipole amplitude. This comes from the fact that different 
radio source populations show up at different flux density limits for different frequencies. Due to this 
effect it seems possible to analyze the structure component of the radio dipole by using different 
frequencies and flux density limits. A kind of tomography of the local universe would be a 
possible application. Radio telescopes like LOFAR, ASKAP or SKA will be ideal to create the necessary 
radio catalogues at different frequencies for this purpose.

\subsection{Line of sight dependence}

It was shown by \citet{Singal11} that the amplitude of a linear dipole estimator (using the NVSS survey), for different areas of the sky, varies like $\cos(\theta)$, where $\theta$ is the angle measured between the line of sight and the dipole direction. This analysis is in agreement with the assumption that the radio dipole is dominated by a kinetic contribution. 

We now discuss whether a radio dipole which is partly due to a contribution from a local void, would be in conflict with this observed behaviour or not. Therefore we investigate how the dipole amplitude from a void varies with respect to the angle $\theta$ (for simplicity we assume a constant density contrast inside the void here). Any prefactors, which do not depend on $\theta$, are not relevant here. 

The effect of a void-like structure on the dipole amplitude is proportional to the length of the line of sight inside the void. If we are inside the void, we need to add the forward and backward contribution, since a dipole estimator also picks up both parts. In order to find this length, we use the equation of a circle (radius $R_{\rm v}$) with an offset of $r_{\rm v}$ in the x-direction
\begin{equation}
 (x-r_{\rm v})^2+y^2=R_{\rm v}^2 \ .
\end{equation}
Using polar coordinates $l,\theta$ with $x=l \cos{\theta},y=l \sin{\theta}$ we can transform this to
\begin{equation}
\label{twosolutions}
 \frac{l}{R_{\rm v}}=\frac{r_{\rm v}}{R_{\rm v}} \cos(\theta) \pm \sqrt{\frac{r^2_{\rm v}}{R^2_{\rm v}} \cos{\theta}^2 +1 -\frac{r^2_{\rm v}}{R^2_{\rm v}}} \ .
\end{equation}
For the case $R_{\rm v} \leq r_{\rm v}$ we subtract one solution from the other to get the distance between the two crossings of the line of sight with the void boundaries. Therefore the dipole signal is proportional to
\begin{equation}
 f = \sqrt{\frac{r^2_{\rm v}}{R^2_{\rm v}} \cos{\theta}^2 +1 -\frac{r^2_{\rm v}}{R^2_{\rm v}}} \ ,
\end{equation}
where $f$ is now the fraction of the dipole effect in the direction $\theta$, meaning $d \propto f$. For the case of an observer on the edge of the void ($r_{\rm v}=R_{\rm v}$), this simplifies to a cosine and therefore behaves exactly like a kinetic dipole.

For the case $R_{\rm v} \geq r_{\rm v}$ one solution of (\ref{twosolutions}) is negative. For an observer inside the void the fraction of the dipole effect depends on the difference of the length of the line of sight in the forward and backward direction. Thus, we need to add up the two solutions of (\ref{twosolutions}) and obtain
\begin{equation}
 f = \cos{\theta} \ .
\end{equation}
Therefore an observer inside a void sees an effect for the dipole estimation which behaves like a cosine. 
So we can not distinguish this case from a kinetic dipole by its angular dependence.

\begin{figure}[!t]
\begin{center}
 \includegraphics[angle=270,width=8cm]{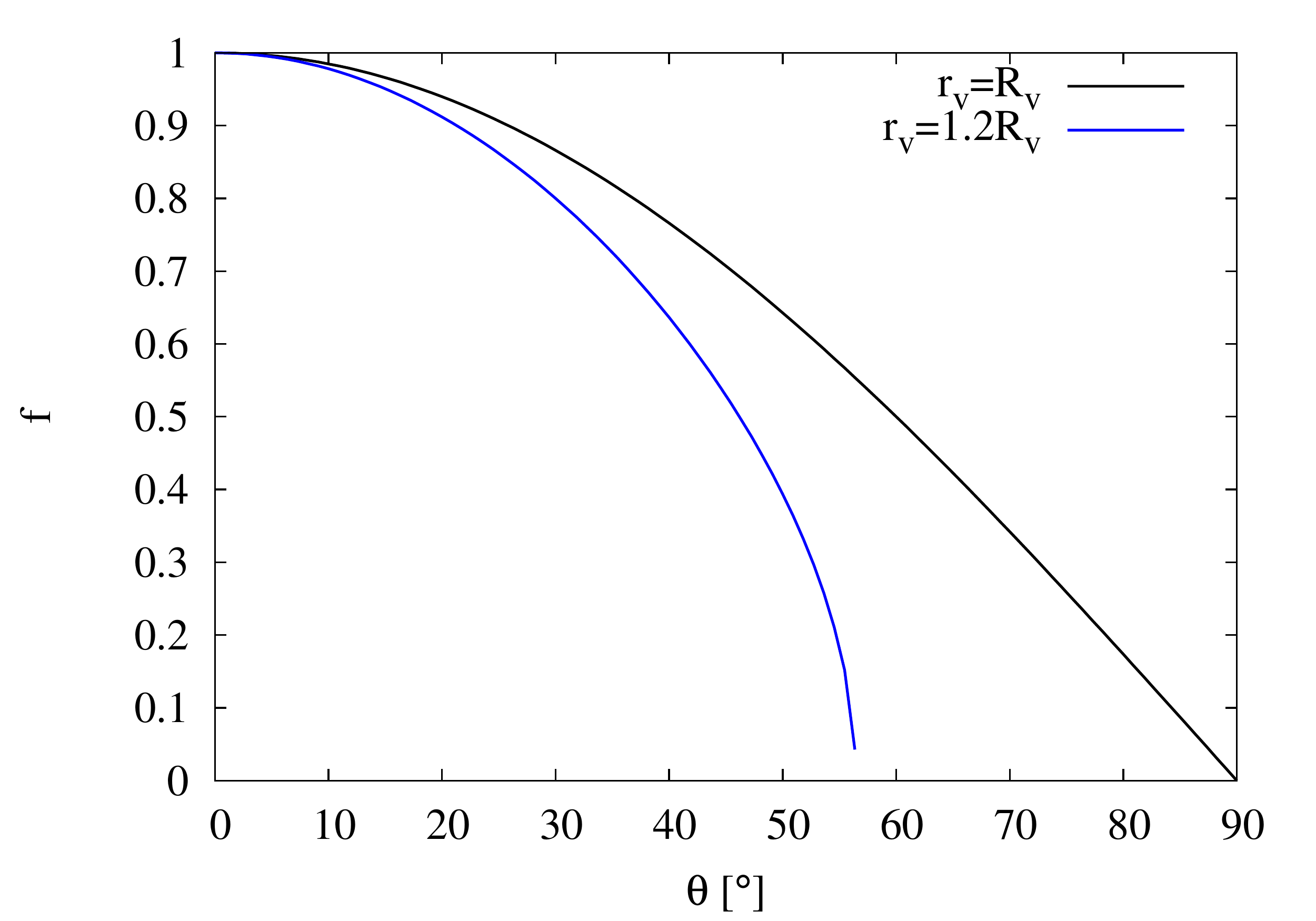} 
\caption{Fraction of dipole effect $f$ versus angle between dipole direction and line of sight $\theta$ for different ratios of the offset vector $r_{\rm v}$ to the void radius $R_{\rm v}$.} \label{angledep}
\end{center}
\end{figure}

In figure \ref{angledep} we show the relation of $f$ versus $\theta$ for different values of 
$r_{\rm v}/R_{\rm v}$. We see that this relation is steeper when the observer is outside the 
void ($r_{\rm v} > R_{\rm v}$). In future surveys, such an analysis can help to separate the kinetic dipole contribution from a structural component and also give an estimation of $r_{\rm v}/R_{\rm v}$.

The case investigated in this work is an observer living close to the edge of a void-like structure. This scenario cannot easily be distinguished from a pure kinetic dipole by this method, since both will behave like a cosine in angular dependence. Therefore the work of \citep{Singal11} is not in conflict to the investigated scenario here.

\section{Conclusion}
\label{Conclusion}

We have been able to develop a model that can describe the influence of spherically symmetric local structures 
on linear dipole estimators. This model was tested and confirmed by computer simulations to a high 
level of accuracy. From this model we learn how the structure parameters (void size $R_{\rm v}$, 
observer distance from the void centre $r_{\rm v}$ and void density contrast $\delta$)
influence the structure dipole amplitude.

Our analytical model requires a constant background density, which is a reasonable approximation 
at small redshifts. In order to include the effects of cosmic expansion and galaxy evolution, the 
dipole contribution for the realistic void model was estimated by means of simulations of the
number counts of radio galaxies.
Not included in this work are estimations of the dipole contribution from 
several galaxy clusters or other structures; because the dipole estimator is linear, these would 
just be a sum of terms similar to the ones calculated in this paper.

One might ask if a void as considered in this work would
show up in the CMB. It is clear that the size of the effect will depend
on the distance of the observer from the centre of the void. The CMB
dipole will be maximally affected for an observer sitting at the edge of
the void.  There is no contribution to the CMB dipole if the observer sits 
in the centre and the effect would show up at much smaller angular scales 
if the observer were far away form the void. A CMB dipole would be
induced by the integrated Sachs-Wolfe effect (or the Rees-Sciama effect if
non-linear effects play a role) and thus the dynamics of
the void profile would be important to determine its amplitude.  
Such structures have been studied previously, e.g. by \citet{ISWVoid1}, \citet{ISWVoid2}, \citet{ISWVoid3}, 
\citet{ISWVoid4}, \citet{ISWVoid5}, \citet{ISWVoid6} and \citet{ISWVoid7}. 
An order of magnitude estimate of the maximum possible effect 
\citep{ISWVoid8} gives
\begin{equation}
\Delta T/T \sim \delta^{3/2} \left(\frac{R_{\rm v}}{R_{\rm H}} \right)^3 \sim 2 \times 10^{-4} 
\end{equation}
for the model considered here. This shows that the void-like structure considered in this paper would not be in tension with
the observed CMB dipole, but might contribute to the CMB anomalies at
small multipole moments. A detailed study of this topic is beyond the
scope of this work.  

For the void model of \cite{Void1} for our local environment, we have run  simulations which include a radio sky 
model from \cite{SSS}. We found  that such a void already has a significant effect on the dipole estimation 
for surveys like the NVSS. The dipole amplitude measured by the linear estimator from 
\citet{firstpaper} of this void is expected to be $\tilde d_{\rm{void}} = 0.21 \pm 0.01 \ \times 10^{-2}$. 
The discrepancy between radio and CMB dipole measurements can be relaxed by 
such a contribution, but the difference cannot be explained completely by the contribution from a single,
 realistic void. In forthcoming surveys, with lower flux density limits, the effect of local 
structure will become even more important.

\begin{acknowledgements} 
MR and DJS acknowledge financial support from the Friedrich Ebert Stiftung and from the Deutsche Forschungsgemeinschaft, grant RTG 1620 ``Models of Gravity''. DB is supported by UK Science 
and Technology Facilities Council, grant ST/K00090X/1. Please contact the authors to request 
access to research materials discussed in this paper. 
\end{acknowledgements}

\bibliographystyle{aa} 
\bibliography{dipole.bib}

\end{document}